\documentclass[apj, iop]{emulateapj}
\shorttitle{Simulations of radio emission from ultracool dwarfs}
\shortauthors{Kuznetsov et al.}
\begin{document}
\title{Comparative analysis of two formation scenarios of bursty radio emission from ultracool dwarfs}
\author{A.A. Kuznetsov\altaffilmark{1,2}, J.G. Doyle\altaffilmark{1}, S. Yu\altaffilmark{1}, G. Hallinan\altaffilmark{3,4}, A. Antonova\altaffilmark{5}, \& A. Golden\altaffilmark{6}}
\altaffiltext{1}{Armagh Observatory, Armagh BT61 9DG, Northern Ireland}
\altaffiltext{2}{Institute of Solar-Terrestrial Physics, Irkutsk 664033, Russia}
\altaffiltext{3}{National Radio Astronomy Observatory, 520 Edgemont Road, Charlottesville, VA 22903, USA}
\altaffiltext{4}{Department of Astronomy, University of California, Berkeley, CA 94720, USA}
\altaffiltext{5}{Department of Astronomy, Faculty of Physics, St. Kliment Ohridski University of Sofia, 1164 Sofia, Bulgaria}
\altaffiltext{6}{Yeshiva University, NY 10461, USA}
\begin{abstract}
Recently, a number of ultracool dwarfs have been found to produce periodic radio bursts with high brightness temperature and polarization degree; the emission properties are similar to the auroral radio emissions of the magnetized planets of the Solar System. We simulate the dynamic spectra of radio emission from ultracool dwarfs. The emission is assumed to be generated due to the electron-cyclotron maser instability. We consider two source models: the emission caused by interaction with a satellite and the emission from a narrow sector of active longitudes; the stellar magnetic field is modeled by a tilted dipole. We have found that for the dwarf TVLM 513--46546, the model of the satellite-induced emission is inconsistent with the observations. On the other hand, the model of emission from an active sector is able to reproduce qualitatively the main features of the radio light curves of this dwarf; the magnetic dipole seems to be highly tilted (by about $60^{\circ}$) with respect to the rotation axis.
\end{abstract}

\section{Introduction}\label{Intro}
Very low-mass stars and brown dwarfs (collectively known as ultracool dwarfs) represent a distinctive class of radio sources. Firstly, they violate the empirical G\"udel-Benz relation between the radio and X-ray emission \citep{ber05, ber06}. For a wide range of objects (including stars of various spectral classes and solar and stellar flares), the spectral radio luminosity $L_{\mathrm{R}}$ and X-ray luminosity $L_{\mathrm{X}}$ satisfy the relation $L_{\mathrm{X}}/L_{\mathrm{R}}\simeq 10^{15.5\pm 0.5}$ Hz \citep{gud93} which indicates the common origin of these emissions. For the stars with spectral class M7 and cooler, the X-ray luminosity abruptly decreases \citep{ber10} as a result of transition from hot ionized coronae to cooler neutral atmospheres \citep{sch09}. Nevertheless, it was recently discovered that a number of very low-mass stars and brown dwarfs in the spectral range M7--L3 are the sources of unexpectedly intense radio emission in the GHz frequency range \citep{ber01, ber08b, ber10, ber02, ber06, bur05, ost06, ost09, pha07, ant08, rav11}; the emission intensity is comparable to or even exceeds that of the early M dwarfs. Secondly, the radio emission from ultracool dwarfs can include (in addition to a quiescent component) short periodic pulses with high brightness temperature and almost 100\% circular polarization, whose period seems to coincide with the stellar rotation period \citep{ber05, ber08a, ber09, hal06, hal07, hal08, doy10}. The total pulse duration can be as short as $\Delta t\lesssim 10^{-2}T_{\mathrm{rot}}$ \citep{hal07, ber08a, ber09}, where $T_{\mathrm{rot}}$ is the stellar rotation period, which suggests high emission directivity and small (relative to the stellar radius) source size. The emission pulses can exhibit fine temporal structure that is repeated each rotation period, with the duration of individual subpulses down to a few seconds \citep{hal07}.

The above features are untypical for the stellar radio emission and more similar to the auroral radio emission of the magnetized planets of the Solar System \citep{zar98}, but with much higher frequency and intensity. In particular, the most likely emission mechanism is the electron-cyclotron maser instability \citep{wu79, mel82, tre06} which produces emission near the local electron cyclotron frequency. In order to provide the observed emission frequencies, ultracool dwarfs must possess magnetic fields of several thousand gauss. Observations of Zeeman broadening of molecular lines in cool stars (with the spectral class up to M9) have confirmed the existence of surface magnetic fields with strengths up to 3900 G \citep{rei07, rei10, rei09}.

The interiors of ultracool dwarfs are expected to be fully convective \citep{cha97}, and therefore the magnetic dynamo processes in these objects should differ significantly from those in the Sun and other solar-type stars; several models have been proposed \citep[e.g.,][]{dob06, cha06, bro08}. The observations have shown that the morphology of the periodic light curves and hence the magnetic field structure remain stable at timescales of up to several months \citep{doy10}, although observations separated by longer times (years) can demonstrate considerable variations \citep{ant07}. Such behaviour can be reconciled with a turbulent dynamo model.

The magnetic field topologies of the radio-emitting ultracool dwarfs are unknown. The planets of the Solar System have, in general, dipole-like magnetic fields with a dipole either nearly parallel to the rotation axis (as for the Earth, Jupiter, and Saturn) or highly tilted with respect to the rotation axis (as for Uranus and Neptune) \citep{rus10}. Spectropolarimetric observations of a sample of M5--M8 dwarfs \citep{mor10} have revealed that these stars tend to have dipole-like axisymmetric magnetic fields (although other topologies have been observed as well). On the other hand, simulations of the $\alpha^2$ dynamo in fully convective stars \citep{kuk99, dob06, cha06} predict approximately dipole-like magnetic field with the dipole perpendicular to the rotation axis. Observations of the cyclotron-maser radio emission offer a unique (and often the only) way to study the magnetic field topology and other parameters of the magnetospheres of ultracool dwarfs; the obtained results, in turn, will be very important for the further development of the stellar dynamo theory.

The aim of this work is to analyze and compare different source models that can result in formation of the short periodic (or quasi-periodic) emission pulses from ultracool dwarfs. Namely, we consider modulation of the emission by the stellar rotation (in a non-axisymmetric magnetosphere model) and the emission induced by a satellite orbiting the ultracool dwarf; both models are similar to the cases observed in the Solar System. The used models are described in Section \ref{Model}. The simulation results are presented in Section \ref{Results} and compared with the observations in Section \ref{Observations}. The conclusions are drawn in Section \ref{Conclusion}.

\begin{figure}
\resizebox{4.2cm}{!}{\includegraphics[clip=true]{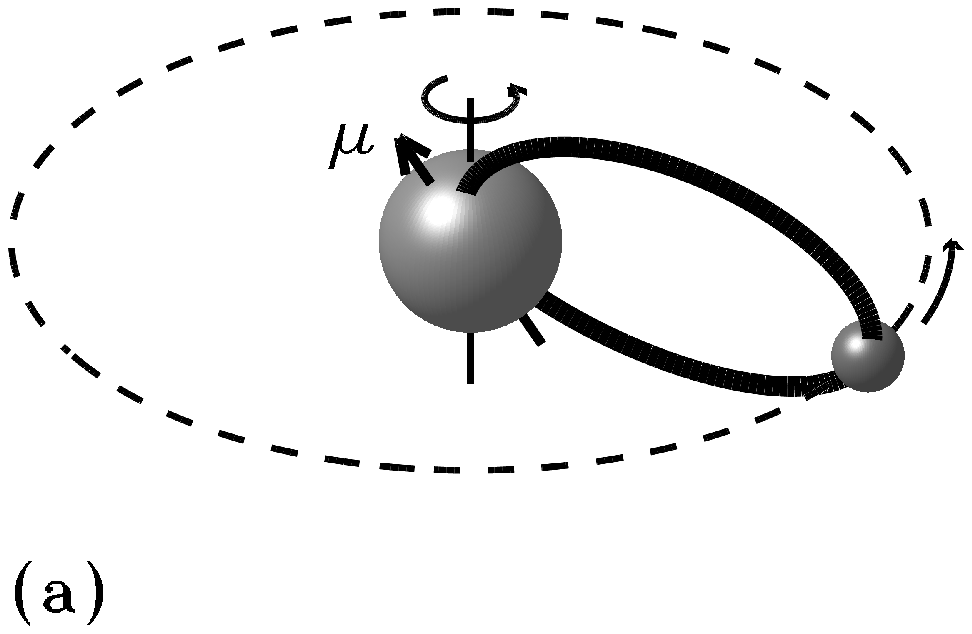}}~
\resizebox{4.2cm}{!}{\includegraphics[clip=true]{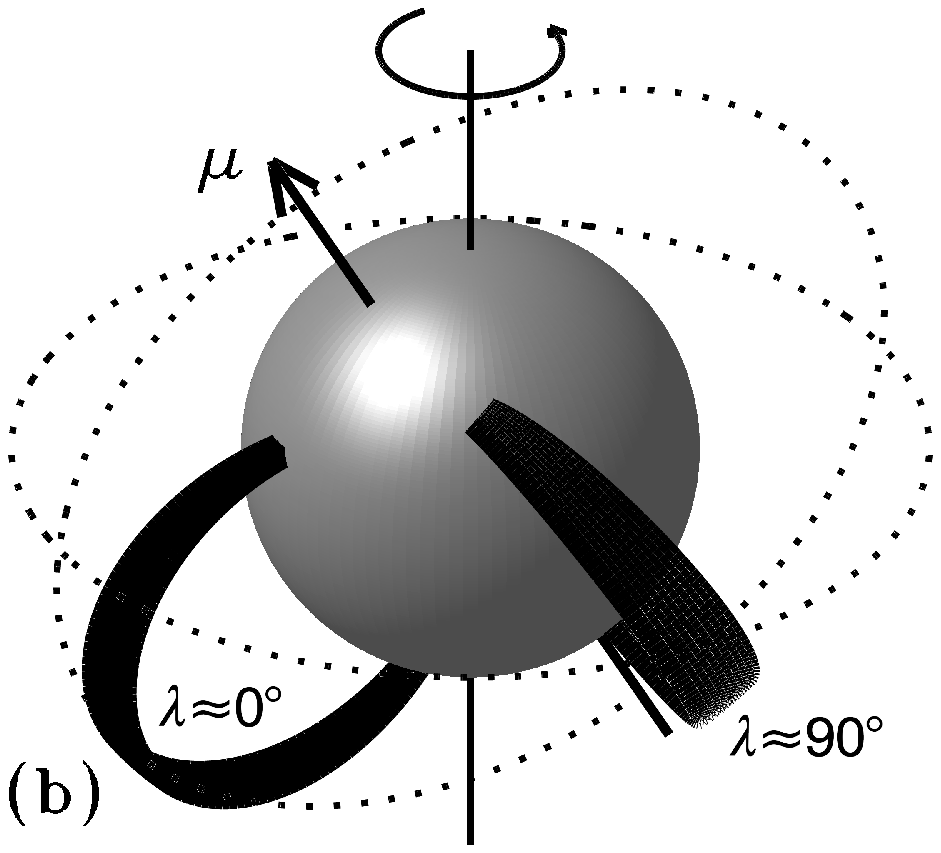}}
\caption{Sketch of the emission source models: a) the model with the star-satellite interaction; b) the model with emission from a set of magnetic field lines (the rotational and magnetic equatorial planes are shown by dotted lines). The magnetic dipole axis is shown by bold arrows.}
\label{FigGeom}
\end{figure}

\section{Model}\label{Model}
We have developed a numerical code to simulate the dynamic spectra of the radio emission. It is assumed that the magnetic field of the ultracool dwarf is dipole-like with the dipole located at the star center (no offset) and tilted by the angle $\delta$ relative to the rotation axis. The rotation axis is inclined relative to the line of sight by the angle $i$. Since the periodic pulses can be very short, their source should occupy a narrow range of longitudes. We therefore consider two models that can provide the required source geometry:
\begin{itemize}
\item
Satellite-induced radio emission (analogous to the Io-Jupiter system). In this model, a satellite (exoplanet) orbits the ultracool dwarf and its interaction with the magnetospheric plasma results in electron acceleration (possibly, due to the excitation of Alfv\'en waves). Thus we assume that the radio emission is generated at the satellite flux tube, i.e., at the magnetic field line passing through the satellite (see Fig. \ref{FigGeom}a). Since the Alfv\'en waves propagate at a finite speed, rapid rotation of the dwarf can shift the radio-emitting region ahead of the satellite; however, this effect is neglected in this study. We also assume that the satellite orbit is circular and the orbital plane coincides with the stellar equatorial plane.
\item
Radio emission from a sector of active longitudes (similar to the case of Jovian hectometric radiation). In this model, the emission is generated at fixed (in the rotating frame of the star) magnetic field lines. Namely, we assume that the ``active'' field lines cross the magnetic equatorial plane at a given distance from the star center (i.e., have the same $L$-shell numbers) and their magnetic longitudes $\lambda$ are confined within a narrow range $\Delta\lambda\ll 1$. Figure \ref{FigGeom}b demonstrates two sets of ``active'' lines: with $\lambda\simeq 0^{\circ}$ and $\lambda\simeq 90^{\circ}$ (the magnetic longitude is calculated relative to the plane containing the rotation axis and the dipole axis). The electrons are accelerated, most likely, due to magnetic reconnection at the boundary of the magnetospheric corotation region. 
\end{itemize}
Similar models (but with significantly different parameters) were studied by \citet{hes11} in application to the exoplanetary or exoplanet-induced stellar radio emission. 

The emission is produced due to the electron-cyclotron maser instability. We consider two types of unstable electron distributions: the loss cone and the shell (or horseshoe) distribution \citep[see, e.g.,][]{tre06}. The shell distributions have been observed in situ within the sources of terrestrial and Saturnian kilometric radiation \citep[e.g.,][]{del98, erg00, lam10}, while the characteristics of Jovian decametric radiation agree better with the predictions for the loss-cone-driven instability \citep{hes07, hes08, mot10}. Thus both types of distributions can occur (possibly, simultaneously) in the magnetospheres of ultracool dwarfs. For the loss-cone-driven instability, the emission frequency $f$ and propagation direction $\theta_0$ (relative to the local magnetic field) are given by \citep{kuz07, hes11}
\begin{equation}\label{fLC}
f\simeq f_{\mathrm{B}}/\sqrt{1-v^2/c^2}>f_{\mathrm{B}},
\end{equation}
\begin{equation}
\cos\theta_0\simeq (v/c)/\cos\alpha_{\mathrm{c}},
\end{equation}
where $f_{\mathrm{B}}$ is the local electron cyclotron frequency, $v$ is the typical speed of the energetic electrons, and $\alpha_{\mathrm{c}}$ is the loss-cone boundary. The emission is assumed to be azimuthally symmetric with respect to the local magnetic field, thus it is beamed along a thin conical sheet with the half-apex angle of $\theta_0$. The loss-cone boundary follows the transverse adiabatic invariant
\begin{equation}
\sin^2\alpha_{\mathrm{c}}=f_{\mathrm{B}}/f_{\mathrm{Bmax}},
\end{equation}
where $f_{\mathrm{Bmax}}$ is the electron cyclotron frequency at the footprint of the radio-emitting magnetic field line (where it enters dense atmospheric layers).

For the shell-driven instability, the emission frequency is given by \citep{hes08, kuz10, hes11}
\begin{equation}\label{fshell}
f\simeq f_{\mathrm{B}}\sqrt{1-v^2/c^2}<f_{\mathrm{B}},
\end{equation}
and the emission direction $\theta_0\simeq 90^{\circ}$ irrespective of the electron energy or the source location.

We compute the dynamic spectra in the following way: for a given time, frequency, and radio-emitting magnetic field line (the line is either fixed or determined by the current satellite position relative to the magnetic dipole), we find the emission source coordinates. Then we calculate the angle $\theta$ between the local magnetic field and the line of sight. Contribution of the considered field line to the emission intensity is calculated as
\begin{equation}\label{gauss}
I(t, f)\sim\exp\left[-(\theta-\theta_0)^2/\Delta\theta^2\right],
\end{equation}
where $\Delta\theta$ is the beam half-width. The radio emission is assumed to be 100\% circularly polarized; the emissions originating from the northern and southern magnetic hemispheres have opposite polarization signs. As Eq. (\ref{gauss}) implies, the considered model does not take into account variations of the emission intensity with height; therefore it provides only some general emission patterns in the time-frequency domain, i.e., where and when the pulses can occur.

Radio emission can propagate only if its frequency exceeds the local plasma frequency, which can result in an additional time-dependent low-frequency cutoff \citep{hes11}. Due to high gravity and low plasma temperature (as indicated by very weak X-ray emission), the plasma density scale height in the magnetospheres of ultracool dwarfs is expected to be much less than the stellar radius. Therefore we consider a simplified occultation effect: the emission (regardless of frequency) is completely absorbed below some critical distance from the star center $R_{\mathrm{cr}}$ and propagates freely above that distance; we assume that $R_{\mathrm{cr}}$ approximately equals the stellar radius $R_*$. The other propagation effects (such as refraction, scattering, and depolarization) are assumed to be negligible.

\section{Simulation results}\label{Results}
Even within the above mentioned simplified models, it is impossible to consider here all the possible combinations of the source parameters. Therefore we restrict ourselves to searching for the cases when the simulated radio emission would be similar to the observed one. As an example, we use the most studied radio-emitting ultracool dwarf TVLM 513--46546 (hereafter TVLM 513). This M9 dwarf has the rotation period of $T_{\mathrm{rot}}=1.96$ h \citep{hal07, doy10}; its mass and radius can be estimated as $M_*=(0.06-0.08)M_{\sun}$ and $R_*=(0.097-0.109)R_{\sun}$, respectively \citep{dah02, hal06}. The observed projected rotation velocity of $v\sin i\simeq 60$ km $\mathrm{s}^{-1}$ \citep{moh03}, together with the mentioned radius and rotation period, imply that the rotation axis should be highly inclined relative to the line of sight ($65^{\circ}\lesssim i\lesssim 90^{\circ}$) \citep{hal06, ber08a}. In the below simulations, we use the following average values of the dwarf mass and radius: $M_*=0.07M_{\sun}$, $R_*\simeq 0.1R_{\sun}=70\,000$ km.

In the simulations, we assume that the radio beam half-width is $\Delta\theta=2^{\circ}$ and the speed of the energetic electrons is $v=0.3c$. The electrons responsible for the planetary radio emissions in the Solar System typically have lower energies ($v\simeq 0.1c$), but we chose a higher value in order to make the differences between the loss-cone-driven and shell-driven instabilities more apparent.

\begin{figure*}
\centerline{\includegraphics{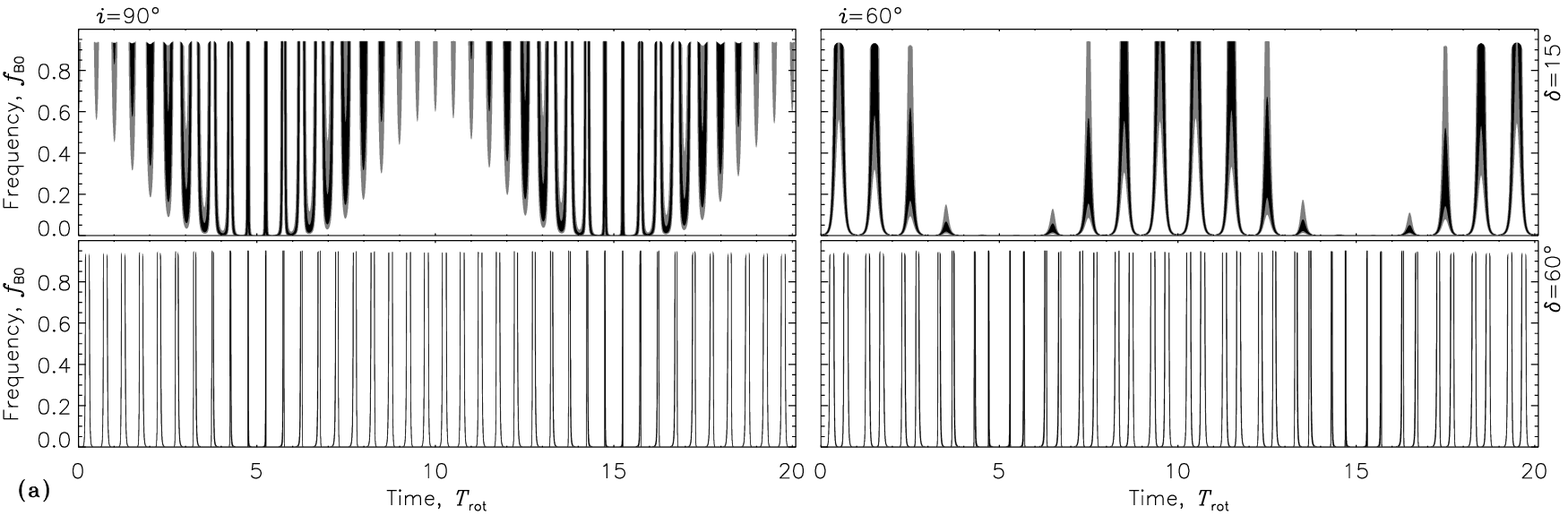}}
\centerline{\includegraphics{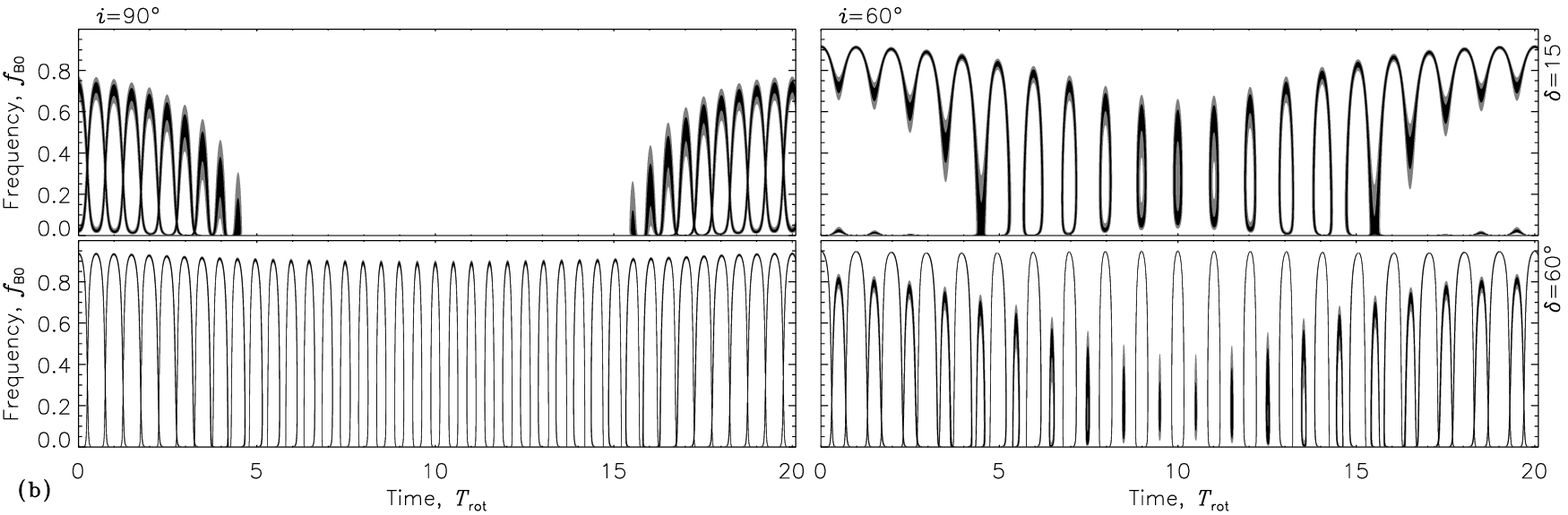}}
\caption{Model dynamic spectra (Stokes $I$) of a satellite-induced radio emission for the different inclinations of the rotation/orbital axis $i$ and magnetic dipole tilts $\delta$. a) Shell-driven emission. b) Loss-cone-driven emission.}
\label{FigSatLarge}
\end{figure*}

\begin{figure}
\resizebox{\hsize}{!}{\includegraphics{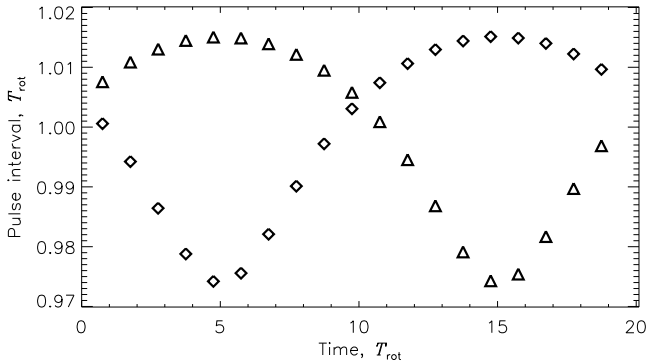}}
\caption{Time intervals between the emission pulses with the same polarization at the fixed frequency of $f=0.6f_{\mathrm{B0}}$ (for the model with a satellite, $i=90^{\circ}$, $\delta=60^{\circ}$, and shell-driven instability). Only the pulses with odd numbers are considered (i.e., approximately one pulse per rotation period). Diamonds and triangles correspond to the pulses with positive ($V>0$) and negative ($V<0$) circular polarizations, respectively; the abscissa represents the centers of the time intervals.}
\label{FigIntervals}
\end{figure}

\begin{figure}
\resizebox{\hsize}{!}{\includegraphics{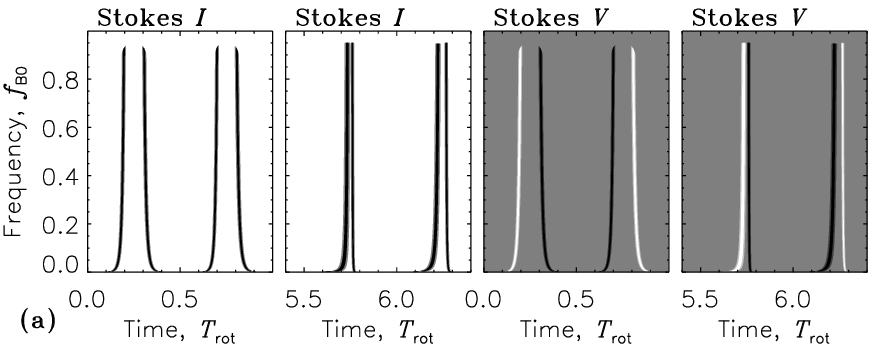}}\\
\resizebox{\hsize}{!}{\includegraphics{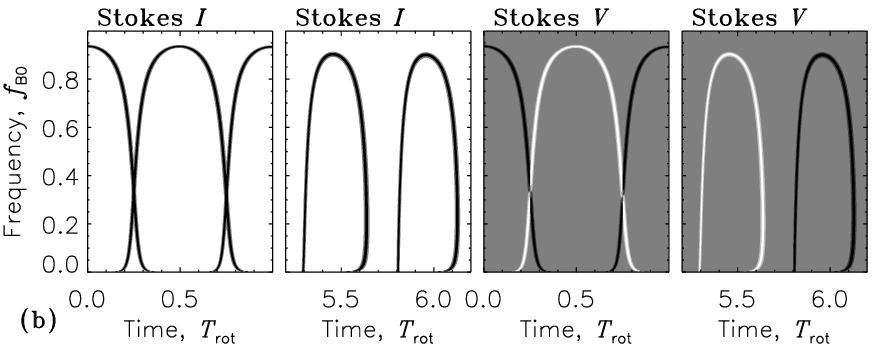}}
\caption{Model dynamic spectra (Stokes $I$ and $V$) of a satellite-induced radio emission for $i=90^{\circ}$ and $\delta=60^{\circ}$; two time intervals corresponding to the different orbital phases of the satellite are shown. The areas which are darker or brighter than the background correspond to the positive or negative values, respectively. a) Shell-driven emission. b) Loss-cone-driven emission.}
\label{FigSatSmall}
\end{figure}

\subsection{Satellite-induced emission}\label{ResSat}
Since we are interested in a periodic radio emission with the period equal or close to the stellar rotation period $T_{\mathrm{rot}}$, we consider the case when the satellite moves relatively slowly, so that its orbital period $T_{\mathrm{orb}}\gg T_{\mathrm{rot}}$. In the below simulations, we assume that $T_{\mathrm{orb}}=20.1T_{\mathrm{rot}}$; for the TVLM 513 system, this corresponds to the orbit radius of $R_{\mathrm{orb}}=24.0R_*$. If the magnetic dipole is axisymmetric, the emission pulse period will equal the satellite orbital period; therefore we consider only the nonzero dipole tilts.

Figure \ref{FigSatLarge} shows the simulated dynamic spectra (emission intensity) for low ($\delta=15^{\circ}$) and high ($\delta=60^{\circ}$) dipole tilts and for two inclinations of the rotation axis ($i=90^{\circ}$ and $60^{\circ}$). The emission frequency in the spectra is expressed in units of the electron cyclotron frequency at the magnetic pole $f_{\mathrm{B0}}$. We have found that if the magnetic dipole tilt is relatively small, then the emission can be detected (i.e., it is beamed towards an observer) only during some favourable orbital phases of the satellite. As a result, the emission pulses at a single frequency occur in series separated by periods of inactivity; the favourable orbital phases are frequency-dependent and depend also on the rotation axis inclination and the characteristics of the electron distribution function.

For the larger dipole tilts, wide-band emission pulses can be detected at any orbital phase of the satellite. The required tilt increases with a decrease of the rotation axis inclination and can be estimated as $\delta\gtrsim 120^{\circ}-i$ (we consider here only the inclinations close to $90^{\circ}$). The dynamic spectra of the shell-driven emission look like series of nearly vertical stripes (with high frequency drift rates); the high-frequency cutoff is determined by the electron cyclotron frequency at the footprint of the satellite flux tube $f_{\mathrm{Bmax}}$ with the relativistic correction (\ref{fshell}). For the loss-cone-driven emission, the dynamic spectra of separate pulses have an ``inverted U'' shape; the high-frequency cutoff is determined mainly by the visibility conditions since the emission frequency and propagation direction are interrelated. At a single frequency we can see, in general, four emission pulses (two with the right circular polarization and two with the left circular polarization) per each rotation period; the pulses with opposite polarizations can overlap thus forming a pulse with a higher intensity and lower polarization degree. If the rotation axis inclination differs from $90^{\circ}$ then the pulses with opposite polarizations become asymmetric (e.g., they have different high-frequency cutoffs); this effect is negligible for the shell-driven emission but significant for the loss-cone-driven emission.

The emission pulses are not strictly periodic; however, each pulse is repeated approximately one rotation period later. Time intervals between the corresponding pulses in the adjacent rotation periods are shown in Fig. \ref{FigIntervals} (these intervals are nearly independent on the source parameters provided that the dipole tilt is sufficiently high). One can see that the time intervals are very close to the stellar rotation period. Variations of the dynamic spectra and light curves from period to period can fall below the time resolution of the instrument if the radius of the satellite orbit is sufficiently large. Therefore, if observations cover only a few consecutive rotation periods, one may conclude that the emission is periodic. However, time intervals between the emission pulses are not constant; moreover, the intervals between the consecutive left-polarized pulses vary in antiphase with the intervals between the right-polarized pulses. Therefore, the dynamic spectra and light curves for a chosen rotation period vary significantly on time scales comparable to the orbital period of the satellite. This can be seen in Fig. \ref{FigSatSmall} which shows enlarged fragments of the dynamic spectra in intensity and polarization for the orbital phases of the satellite near $0^{\circ}$ and $90^{\circ}$ (relative to the line of sight).

\begin{figure*}
\centerline{\includegraphics{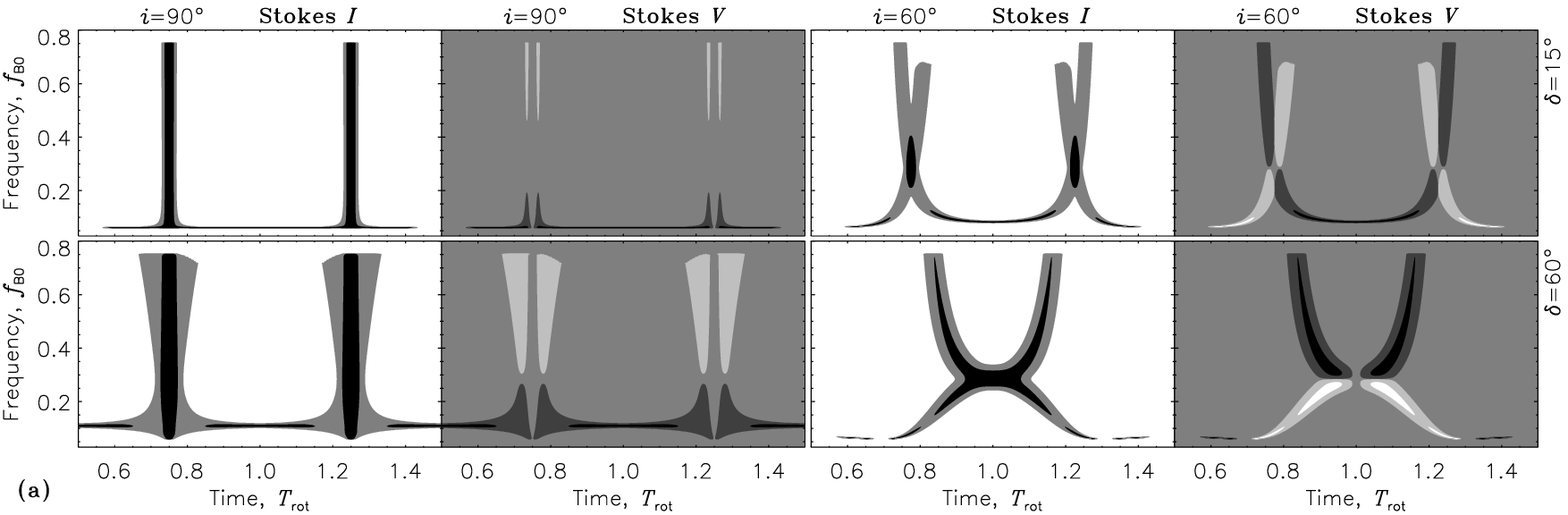}}
\centerline{\includegraphics{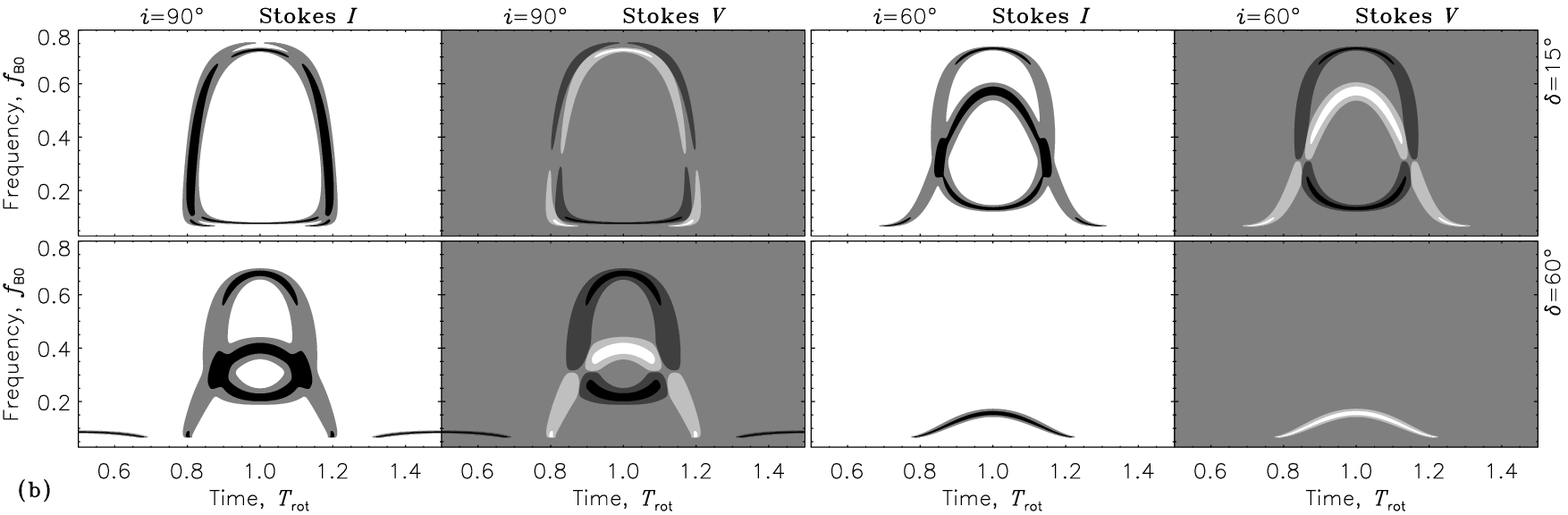}}
\caption{Model dynamic spectra (Stokes $I$ and $V$) of the radio emission from an active sector for the different inclinations of the rotation axis $i$ and magnetic dipole tilts $\delta$. The areas which are darker or brighter than the background correspond to the positive or negative values, respectively. The ``active'' field lines have the magnetic longitudes around $\lambda\simeq 0^{\circ}$. a) Shell-driven emission. b) Loss-cone-driven emission.}
\label{FigARcntr}
\end{figure*}

\begin{figure*}
\centerline{\includegraphics{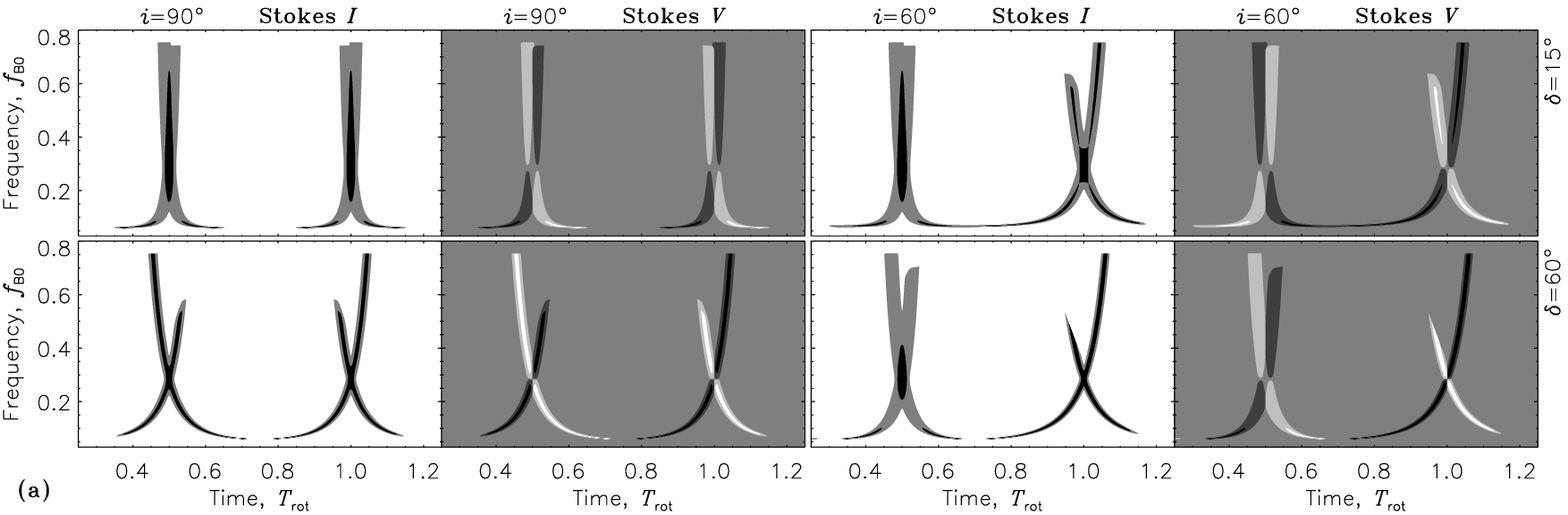}}
\centerline{\includegraphics{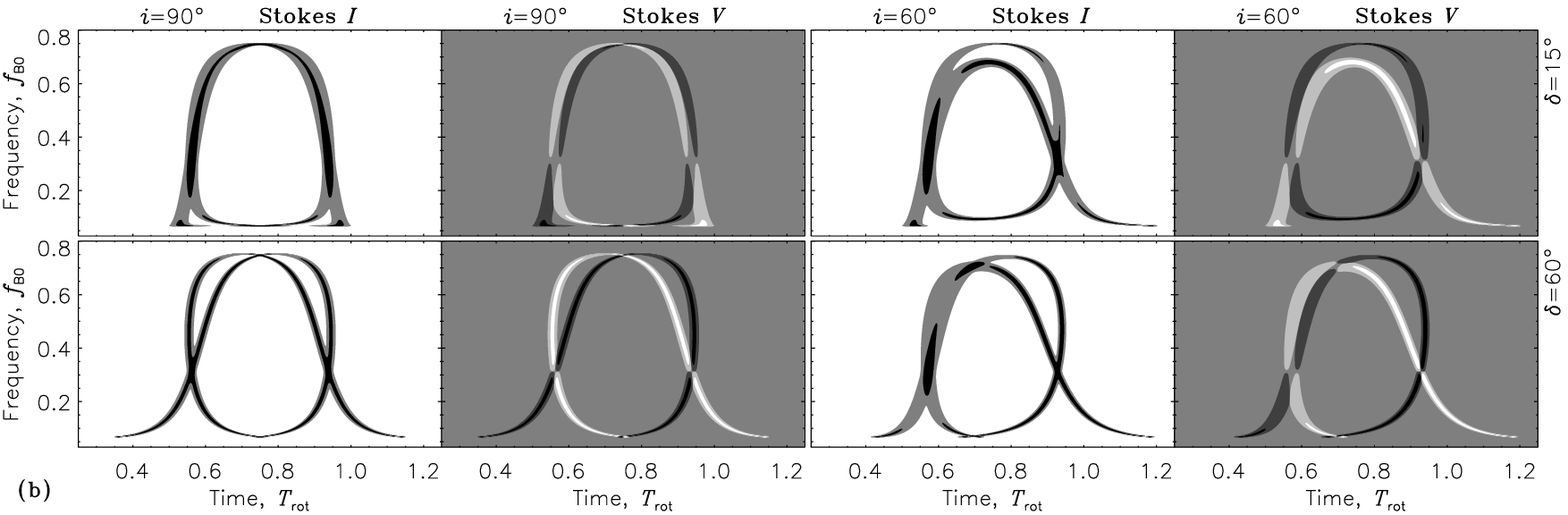}}
\caption{Same as in Fig. \protect\ref{FigARcntr}, for the ``active'' field lines with the magnetic longitudes around $\lambda\simeq 90^{\circ}$.}
\label{FigARside}
\end{figure*}

\subsection{Emission from an active sector}
In this model, the emission is always periodic with the period equal to the stellar rotation period. The emission is generated at the magnetic field lines connected to the particle acceleration region that, in turn, is located in the equatorial plane at (or near) the magnetospheric corotation radius \citep{and88, lin92}. Due to fast rotation, the corotation radii and hence the $L$-shell numbers of the radio-emitting field lines for ultracool dwarfs are expected to be rather small; \citet{rav11} estimated the corotation radius for the radio-emitting M8 dwarf DENIS 1048--3956 as $R_{\mathrm{C}}\simeq (1.2-2.9)R_*$. In the below simulations, we assume that the ``active'' magnetic field lines have $L=2$. We also assume that the angular width of the active sector is $\Delta\lambda=10^{\circ}$; the total emission is calculated as the sum of emissions from ten field lines evenly distributed within the range $\Delta\lambda$.

We should note that the existence of a narrow sector of active longitudes implies that the magnetospheric topology is more complicated than the dipolar model described in Section \ref{Model}; e.g., the dipole can be offset from the rotation axis (like at Neptune) and/or the higher-order components of the magnetic field can be important. Nevertheless, we assume that the magnetic field is still approximately dipole-like, but the conditions favourable for the emission generation are met only at certain field lines.

Figure \ref{FigARcntr} shows the simulated dynamic spectra (emission intensity and polarization) for low ($\delta=15^{\circ}$) and high ($\delta=60^{\circ}$) dipole tilts and for two inclinations of the rotation axis ($i=90^{\circ}$ and $60^{\circ}$). The active sector is located around the magnetic longitude $\lambda_0=0^{\circ}$, i.e., the ``active'' field lines are parallel to the local meridian (see Fig. \ref{FigGeom}b). Since the emission source is symmetric with respect to the plane containing the rotation axis and the dipole axis, the dynamic spectra are also symmetric and the number of emission pulses (at a given frequency) per rotation period is always even. For large rotation axis inclinations ($i\simeq 90^{\circ}$) and low dipole tilts ($\delta\lesssim 15^{\circ}$), the emission sources located in opposite magnetic hemispheres are observed nearly simultaneously which results in weakly-polarized emission. With an increase in the dipole tilt, the emission pulses produced in opposite hemispheres become asymmetric and separated in frequency (so that only the left- or right-polarized emission is observed in certain frequency ranges). A decrease in the rotation axis inclination has a similar effect, but also introduces a separation in time. The loss cone produces the ``inverted-U''-shaped and ring-shaped structures in the dynamic spectrum, while the shell distribution produces either vertical stripes (for $i=90^{\circ}$) or the pulses with a frequency drift (for other inclinations). For high dipole tilts and small rotation axis inclinations, only the emission from one of the magnetic hemispheres can be observed (this effect is more pronounced for the loss-cone-driven emission).

Figure \ref{FigARside} shows the simulated dynamic spectra for the case when the active sector is located around the magnetic longitude $\lambda_0=90^{\circ}$ (see Fig. \ref{FigGeom}b). The magnetic field lines with $\lambda=90^{\circ}$ (or $270^{\circ}$) cross the stellar equatorial plane at the maximal distance from the rotation axis (among the lines with a given $L$), which is expected to result in the most efficient particle acceleration. Like in the previous case, the loss cone produces the ``inverted-U''-shaped and ring-shaped pulses that exhibit a reversal of the frequency drift at the maximal frequency, while the shell distribution produces pulses with a simpler structure, whose high-frequency cutoff is determined by the maximal possible emission frequency at the ``active'' field line and by the occultation effects. At a given frequency, we can observe up to four pulses (two with the right circular polarization and two with the left circular polarization) per rotation period. For the loss-cone-driven emission, the number of emission pulses at a given frequency is always even; for the shell-driven emission, the occultation effects can remove some pulses thus resulting in one or three pulses per rotation period. Such asymmetric light curves can occur if the rotation axis is not perpendicular to the line of sight ($i<90^{\circ}$). Both an increase in the dipole tilt and a decrease in the rotation axis inclination result in an increasing time separation between the left- and right-polarized pulses, but these factors have different effects on the light curves; e.g., if $\delta>(90^{\circ}-i)$, the emission pulses at a single frequency have a ``left--right--left--right'' pattern, while for $\delta<(90^{\circ}-i)$ we will obtain a ``left--right--right--left'' pattern.

\begin{figure}
\resizebox{\hsize}{!}{\includegraphics{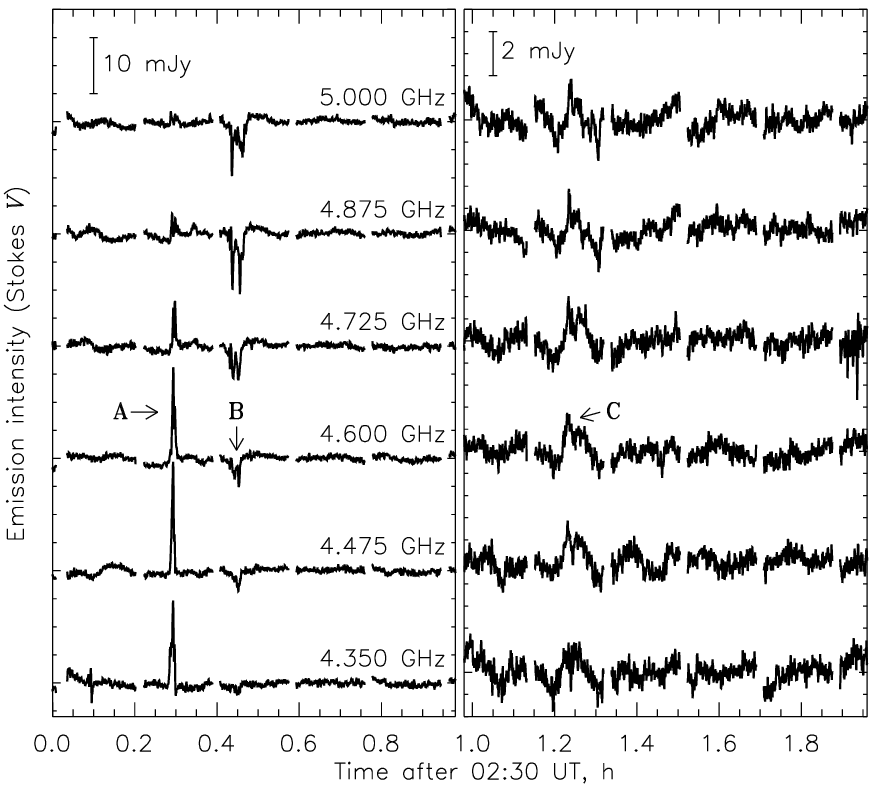}}
\caption{Time profiles of the radio emission (Stokes $V$) from TVLM 513 observed on 2008 May 19. The intensity scales in the first and second halves of the considered time interval are different in order to make the pulse C more visible. The gaps in the time profiles correspond to the time intervals when the instrument calibration was performed.}
\label{FigObs}
\end{figure}

\begin{figure}
\resizebox{\hsize}{!}{\includegraphics{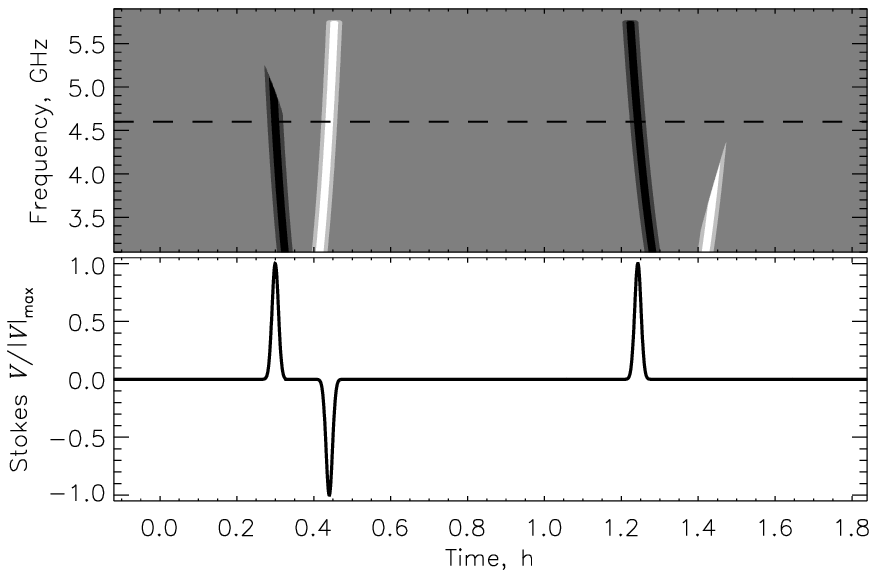}}
\caption{Top: model dynamic spectra (Stokes $V$) of the radio emission from an active sector. Bottom: the corresponding light curve at the frequency of 4.6 GHz. The sumulation parameters are given in the text; they were chosen so as to provide a qualitative agreement of the simulated light curve with the observations.}
\label{FigMod}
\end{figure}

\section{Comparison with observations}\label{Observations}
Most observations of the radio emission from ultracool dwarfs have been performed at one or two frequencies. Only a few dynamic spectra (with a rather narrow frequency range) have been recorded to date. Therefore we focus here on the characteristics of the emission light curves at a single frequency.

TVLM 513 has the longest total duration of radio observations among all ultracool dwarfs, including continuous observation sessions covering several consequent rotation periods. E.g., \citet{hal07} presented the observations that lasted for $\sim 10$ h or about five rotation periods (see Fig. 1 in the cited work for the light curves); the conclusion about a stable period was made. We have re-examined those observations and found that the relative deviation of a pulse period from the average value does not exceed $5\times 10^{-3}$ (for both polarizations). Even more strict limitation on the possible period variations was obtained by \citet{doy10}, who analyzed two data sets (each lasting for $\sim 8$ h) separated by $\sim 40$ days; the period was found to be stable with an accuracy of up to $10^{-5}$. These results cannot be reconciled with the model of satellite-induced emission unless the satellite orbital motion is very slow. However, according to our simulations, to obtain the relative variations of the pulse periods of $\lesssim 5\times 10^{-3}$ one needs the satellite orbital period of $T_{\mathrm{orb}}/T_{\mathrm{rot}}\gtrsim 100$ and thus the orbit radius of $R_{\mathrm{orb}}/R_*\gtrsim 70$. It is highly unlikely that so distant satellite would be able to modulate the stellar radio emission, and therefore the model of satellite-induced emission can be ruled out for TVLM 513. Nevertheless, this model can work at other ultracool dwarfs.

It is interesting to note that \citet{for09} suggested the detection of a companion of TVLM 513 at a distance of about $20R_*$ from the star (i.e., with the parameters close to those considered in Section \ref{ResSat}). However, it seems that this companion does not affect the periodic cyclotron-maser radio emission from the dwarf (at least, it does not affect the pulse occurrence times). Most likely, this is because the satellite, if present, is located beyond the corotation radius of the stellar magnetic field.

Figure \ref{FigObs} presents an example of time profiles of the radio emission from TVLM 513 at different frequencies; due to technical reasons, only the Stokes $V$ radio flux is shown (see \citealt{hal09} for a more complete discussion of the observational data). The emission was observed with the Wide-band Arecibo Pulsar Processor at the Arecibo Observatory\footnote{The Arecibo Observatory is part of the National Astronomy and Ionosphere Center, which is operated by Cornell University under a cooperative agreement with the National Science Foundation.} on 2008 May 19 \citep[see also][]{ant10a, ant10b}. The time interval shown corresponds to one stellar rotation period ($1.96$ h). One can see three emission pulses (A, B, and C), with the pulses A and C being positively polarized ($V>0$) and the pulse B being negatively polarized ($V<0$). Time intervals between the pulses are $t_{\mathrm{B}}-t_{\mathrm{A}}\simeq 0.075T_{\mathrm{rot}}$ and $t_{\mathrm{C}}-t_{\mathrm{A}}\simeq 0.48T_{\mathrm{rot}}$, i.e., it is likely that the pulses A and C are produced in the same source due to the shell-driven instability. The same morphology of the light curves was observed in each rotation period over three consecutive nights (2008 May 18, 19, and 20), although the pulse amplitudes and durations could vary significantly (e.g., the pulse C in Fig. \ref{FigObs} is much weaker than the pulses A and B, but an opposite relation was observed in some periods). A stable temporal structure with a pronounced periodicity implies that the emission source is fixed in the rotating frame of the star.

The facts that there is an odd number of emission pulses per rotation period and the time interval between the pulses with the same polarization sign is close to $T_{\mathrm{rot}}/2$ suggest that (i) the emission comes from an active sector rotated with respect to the plane containing the rotation axis and the dipole axis (e.g., with $\lambda\simeq 90^{\circ}$ or $270^{\circ}$), (ii) the emission is produced due to the shell-driven instability, and (iii) the rotation axis inclination $i$ differs from $90^{\circ}$. As a result, the counterpart of the pulse B is invisible because the emission source is occulted by the star at that time.

Figure \ref{FigMod} shows the simulated dynamic spectrum and light curve for the following simulation parameters: magnetic field strength at the magnetic pole $B_0=2560$ G (which corresponds to the electron cyclotron frequency of $f_{\mathrm{B0}}=7.2$ GHz), rotation axis inclination $i=70^{\circ}$, magnetic dipole tilt $\delta=60^{\circ}$, ``active'' magnetic line number $L=2.15$, active sector longitude $\lambda_0=270^{\circ}$, active sector width $\Delta\lambda=1^{\circ}$, and speed of the energetic electrons (with the shell distribution) $v=0.1c$. The parameters are chosen so as to provide an agreement with the observations. Namely, our aim was to reproduce a three-pulse structure in a sufficiently wide frequency range as well as the above mentioned time intervals between the pulses. Due to the large number of free parameters, the solution presented in Fig. \ref{FigMod} is not unique. However, we have found that only relatively high dipole tilts ($\delta\simeq 50^{\circ}-70^{\circ}$) can be reconciled with the observations. Also, we have estimated the favorable range of the ``active'' magnetic line numbers as $L\simeq 2.0-2.6$, which corresponds to the surface magnetic field values of $B_0\simeq 2800-2100$ G.

Despite a qualitative agreement, there are apparent differences between the simulations and the observations. Firstly, the observed pulses exhibit much higher frequency drift rates than predicted by the simulations. Possibly, this is because either the magnetic field is not exactly dipole-like or the emission sources do not follow the magnetic field lines (e.g., due to the peculiarities of the particle acceleration and precipitation processes). Secondly, the pulses often exhibit fine spectral structure that varies from period to period. Such fine structures are typical of the planetary radio emissions and are caused, most likely, by small-scale irregular plasma inhomogeneities (turbulence) in the radio emission sources. Therefore, dynamic spectra covering both wide frequency range and long time intervals are necessary to recover reliably the magnetic field topology.

It is interesting to note that the pulse C in Fig. \ref{FigObs} has a much lower intensity but a much longer duration than pulse A, so that the total radiated energy is approximately the same. If we assume that these pulses are produced in the same source then the variations of the pulse intensity and duration can be caused by the radiation scattering during propagation, which is different for the different source orientations. An alternative explanation is that the active sector width $\Delta\lambda$ can vary significantly over timescales of about one hour, while the total number of the energetic electrons in this sector remains nearly constant.

\section{Conclusion}\label{Conclusion}
We have simulated the dynamic spectra of radio emission from ultracool dwarfs. The stellar magnetic field was modeled by a tilted dipole. Two source models were considered: the model in which the emission was produced as a result of the star-satellite interaction, and the model with emission from a narrow sector of magnetic longitudes. The emission was assumed to be generated due to the electron-cyclotron maser instability; two unstable electron distributions (the loss cone and the shell distribution) were considered. We have compared the simulation results with the observations of the ultracool dwarf TVLM 513. We have found that:
\begin{itemize}
\item
The model of the satellite-induced emission is inconsistent with the observations since it predicts variable time intervals between the emission pulses, whereas the radio emission from TVLM 513 exhibits a stable period.
\item
The model of emission from an active sector is able to reproduce the main features of the radio light curves of TVLM 513 (such as the pulse polarizations and occurrence times). It seems that the magnetic dipole is highly tilted (by about $60^{\circ}$) with respect to the rotation axis, i.e., the magnetic field topology of the dwarf is similar to that of Uranus. Also, the model with the shell-like (or horseshoe-like) electron distribution fits the observations better than the model with the loss cone.
\item
On the other hand, the considered model of emission from an active sector seems to be oversimplified since it cannot account for, e.g., the observed frequency drift rates, intensity variations with frequency, and variations of the pulse intensity and duration from period to period.
\end{itemize}

The experimental and theoretical studies of radio emission from ultracool dwarfs are only at their beginning. Dynamic spectra of the emission (with high spectral and temporal resolution) covering wide frequency range and long time intervals are necessary to make more definite conclusions about the magnetic field topology, particle acceleration mechanism, and other parameters of the stellar magnetospheres.

\acknowledgements
A.A. Kuznetsov, J.G. Doyle, and S. Yu thank the Leverhulme Trust for financial support. G. Hallinan is a Jansky Fellow of the National Radio Astronomy Observatory. A. Antonova gratefully acknowledges the support of the Bulgarian National Science Fund (grant No. DDVU02/40-2010).

\bibliographystyle{apj}

\begin{thebibliography}{52}
\expandafter\ifx\csname natexlab\endcsname\relax\def\natexlab#1{#1}\fi

\bibitem[{{Andre} {et~al.}(1988){Andre}, {Montmerle}, {Feigelson}, {Stine}, \&
  {Klein}}]{and88}
{Andre}, P., {Montmerle}, T., {Feigelson}, E.~D., {Stine}, P.~C., \& {Klein},
  K.-L. 1988, \apj, 335, 940

\bibitem[{{Antonova} {et~al.}(2008){Antonova}, {Doyle}, {Hallinan}, {Bourke},
  \& {Golden}}]{ant08}
{Antonova}, A., {Doyle}, J.~G., {Hallinan}, G., {Bourke}, S., \& {Golden}, A.
  2008, \aap, 487, 317

\bibitem[{{Antonova} {et~al.}(2010{\natexlab{a}}){Antonova}, {Doyle},
  {Hallinan}, {Golden}, \& {Bourke}}]{ant10a}
{Antonova}, A., {Doyle}, J.~G., {Hallinan}, G., {Golden}, A., \& {Bourke}, S.
  2010{\natexlab{a}}, Bulgarian Astronomical Journal, 14, 58

\bibitem[{{Antonova} {et~al.}(2007){Antonova}, {Doyle}, {Hallinan}, {Golden},
  \& {Koen}}]{ant07}
{Antonova}, A., {Doyle}, J.~G., {Hallinan}, G., {Golden}, A., \& {Koen}, C.
  2007, \aap, 472, 257

\bibitem[{{Antonova} {et~al.}(2010{\natexlab{b}}){Antonova}, {Hallinan},
  {Doyle}, \& {Golden}}]{ant10b}
{Antonova}, A., {Hallinan}, G., {Doyle}, J.~G., \& {Golden}, A.
  2010{\natexlab{b}}, Publications de l'Observatoire Astronomique de Beograd,
  90, 117

\bibitem[{{Berger}(2002)}]{ber02}
{Berger}, E. 2002, \apj, 572, 503

\bibitem[{{Berger}(2006)}]{ber06}
---. 2006, \apj, 648, 629

\bibitem[{{Berger} {et~al.}(2001){Berger}, {Ball}, {Becker}, {Clarke}, {Frail},
  {Fukuda}, {Hoffman}, {Mellon}, {Momjian}, {Murphy}, {Teng}, {Woodruff},
  {Zauderer}, \& {Zavala}}]{ber01}
{Berger}, E., {Ball}, S., {Becker}, K.~M., {Clarke}, M., {Frail}, D.~A.,
  {Fukuda}, T.~A., {Hoffman}, I.~M., {Mellon}, R., {Momjian}, E., {Murphy},
  N.~W., {Teng}, S.~H., {Woodruff}, T., {Zauderer}, B.~A., \& {Zavala}, R.~T.
  2001, \nat, 410, 338

\bibitem[{{Berger} {et~al.}(2010){Berger}, {Basri}, {Fleming}, {Giampapa},
  {Gizis}, {Liebert}, {Mart{\'{\i}}n}, {Phan-Bao}, \& {Rutledge}}]{ber10}
{Berger}, E., {Basri}, G., {Fleming}, T.~A., {Giampapa}, M.~S., {Gizis}, J.~E.,
  {Liebert}, J., {Mart{\'{\i}}n}, E., {Phan-Bao}, N., \& {Rutledge}, R.~E.
  2010, \apj, 709, 332

\bibitem[{{Berger} {et~al.}(2008{\natexlab{a}}){Berger}, {Basri}, {Gizis},
  {Giampapa}, {Rutledge}, {Liebert}, {Mart{\'{\i}}n}, {Fleming}, {Johns-Krull},
  {Phan-Bao}, \& {Sherry}}]{ber08b}
{Berger}, E., {Basri}, G., {Gizis}, J.~E., {Giampapa}, M.~S., {Rutledge},
  R.~E., {Liebert}, J., {Mart{\'{\i}}n}, E., {Fleming}, T.~A., {Johns-Krull},
  C.~M., {Phan-Bao}, N., \& {Sherry}, W.~H. 2008{\natexlab{a}}, \apj, 676, 1307

\bibitem[{{Berger} {et~al.}(2008{\natexlab{b}}){Berger}, {Gizis}, {Giampapa},
  {Rutledge}, {Liebert}, {Mart{\'{\i}}n}, {Basri}, {Fleming}, {Johns-Krull},
  {Phan-Bao}, \& {Sherry}}]{ber08a}
{Berger}, E., {Gizis}, J.~E., {Giampapa}, M.~S., {Rutledge}, R.~E., {Liebert},
  J., {Mart{\'{\i}}n}, E., {Basri}, G., {Fleming}, T.~A., {Johns-Krull}, C.~M.,
  {Phan-Bao}, N., \& {Sherry}, W.~H. 2008{\natexlab{b}}, \apj, 673, 1080

\bibitem[{{Berger} {et~al.}(2009){Berger}, {Rutledge}, {Phan-Bao}, {Basri},
  {Giampapa}, {Gizis}, {Liebert}, {Mart{\'{\i}}n}, \& {Fleming}}]{ber09}
{Berger}, E., {Rutledge}, R.~E., {Phan-Bao}, N., {Basri}, G., {Giampapa},
  M.~S., {Gizis}, J.~E., {Liebert}, J., {Mart{\'{\i}}n}, E., \& {Fleming},
  T.~A. 2009, \apj, 695, 310

\bibitem[{{Berger} {et~al.}(2005){Berger}, {Rutledge}, {Reid}, {Bildsten},
  {Gizis}, {Liebert}, {Mart{\'{\i}}n}, {Basri}, {Jayawardhana}, {Brandeker},
  {Fleming}, {Johns-Krull}, {Giampapa}, {Hawley}, \& {Schmitt}}]{ber05}
{Berger}, E., {Rutledge}, R.~E., {Reid}, I.~N., {Bildsten}, L., {Gizis}, J.~E.,
  {Liebert}, J., {Mart{\'{\i}}n}, E., {Basri}, G., {Jayawardhana}, R.,
  {Brandeker}, A., {Fleming}, T.~A., {Johns-Krull}, C.~M., {Giampapa}, M.~S.,
  {Hawley}, S.~L., \& {Schmitt}, J.~H.~M.~M. 2005, \apj, 627, 960

\bibitem[{{Browning}(2008)}]{bro08}
{Browning}, M.~K. 2008, \apj, 676, 1262

\bibitem[{{Burgasser} \& {Putman}(2005)}]{bur05}
{Burgasser}, A.~J. \& {Putman}, M.~E. 2005, \apj, 626, 486

\bibitem[{{Chabrier} \& {Baraffe}(1997)}]{cha97}
{Chabrier}, G. \& {Baraffe}, I. 1997, \aap, 327, 1039

\bibitem[{{Chabrier} \& {K{\"u}ker}(2006)}]{cha06}
{Chabrier}, G. \& {K{\"u}ker}, M. 2006, \aap, 446, 1027

\bibitem[{{Dahn} {et~al.}(2002){Dahn}, {Harris}, {Vrba}, {Guetter}, {Canzian},
  {Henden}, {Levine}, {Luginbuhl}, {Monet}, {Monet}, {Pier}, {Stone}, {Walker},
  {Burgasser}, {Gizis}, {Kirkpatrick}, {Liebert}, \& {Reid}}]{dah02}
{Dahn}, C.~C., {Harris}, H.~C., {Vrba}, F.~J., {Guetter}, H.~H., {Canzian}, B.,
  {Henden}, A.~A., {Levine}, S.~E., {Luginbuhl}, C.~B., {Monet}, A.~K.~B.,
  {Monet}, D.~G., {Pier}, J.~R., {Stone}, R.~C., {Walker}, R.~L., {Burgasser},
  A.~J., {Gizis}, J.~E., {Kirkpatrick}, J.~D., {Liebert}, J., \& {Reid}, I.~N.
  2002, \aj, 124, 1170

\bibitem[{{Delory} {et~al.}(1998){Delory}, {Ergun}, {Carlson}, {Muschietti},
  {Chaston}, {Peria}, {McFadden}, \& {Strangeway}}]{del98}
{Delory}, G.~T., {Ergun}, R.~E., {Carlson}, C.~W., {Muschietti}, L., {Chaston},
  C.~C., {Peria}, W., {McFadden}, J.~P., \& {Strangeway}, R. 1998, \grl, 25,
  2069

\bibitem[{{Dobler} {et~al.}(2006){Dobler}, {Stix}, \& {Brandenburg}}]{dob06}
{Dobler}, W., {Stix}, M., \& {Brandenburg}, A. 2006, \apj, 638, 336

\bibitem[{{Doyle} {et~al.}(2010){Doyle}, {Antonova}, {Marsh}, {Hallinan}, {Yu},
  \& {Golden}}]{doy10}
{Doyle}, J.~G., {Antonova}, A., {Marsh}, M.~S., {Hallinan}, G., {Yu}, S., \&
  {Golden}, A. 2010, \aap, 524, A15

\bibitem[{{Ergun} {et~al.}(2000){Ergun}, {Carlson}, {McFadden}, {Delory},
  {Strangeway}, \& {Pritchett}}]{erg00}
{Ergun}, R.~E., {Carlson}, C.~W., {McFadden}, J.~P., {Delory}, G.~T.,
  {Strangeway}, R.~J., \& {Pritchett}, P.~L. 2000, \apj, 538, 456

\bibitem[{{Forbrich} \& {Berger}(2009)}]{for09}
{Forbrich}, J. \& {Berger}, E. 2009, \apjl, 706, L205

\bibitem[{{G\"udel} \& {Benz}(1993)}]{gud93}
{G\"udel}, M. \& {Benz}, A.~O. 1993, \apjl, 405, L63

\bibitem[{{Hallinan}(2009)}]{hal09}
{Hallinan}, G. 2009, PhD thesis, NUI Galway

\bibitem[{{Hallinan} {et~al.}(2006){Hallinan}, {Antonova}, {Doyle}, {Bourke},
  {Brisken}, \& {Golden}}]{hal06}
{Hallinan}, G., {Antonova}, A., {Doyle}, J.~G., {Bourke}, S., {Brisken}, W.~F.,
  \& {Golden}, A. 2006, \apj, 653, 690

\bibitem[{{Hallinan} {et~al.}(2008){Hallinan}, {Antonova}, {Doyle}, {Bourke},
  {Lane}, \& {Golden}}]{hal08}
{Hallinan}, G., {Antonova}, A., {Doyle}, J.~G., {Bourke}, S., {Lane}, C., \&
  {Golden}, A. 2008, \apj, 684, 644

\bibitem[{{Hallinan} {et~al.}(2007){Hallinan}, {Bourke}, {Lane}, {Antonova},
  {Zavala}, {Brisken}, {Boyle}, {Vrba}, {Doyle}, \& {Golden}}]{hal07}
{Hallinan}, G., {Bourke}, S., {Lane}, C., {Antonova}, A., {Zavala}, R.~T.,
  {Brisken}, W.~F., {Boyle}, R.~P., {Vrba}, F.~J., {Doyle}, J.~G., \& {Golden},
  A. 2007, \apjl, 663, L25

\bibitem[{{Hess} {et~al.}(2008){Hess}, {Cecconi}, \& {Zarka}}]{hes08}
{Hess}, S., {Cecconi}, B., \& {Zarka}, P. 2008, \grl, 35, L13107

\bibitem[{{Hess} {et~al.}(2007){Hess}, {Mottez}, \& {Zarka}}]{hes07}
{Hess}, S., {Mottez}, F., \& {Zarka}, P. 2007, Journal of Geophysical Research
  (Space Physics), 112, A11212

\bibitem[{{Hess} \& {Zarka}(2011)}]{hes11}
{Hess}, S.~L.~G. \& {Zarka}, P. 2011, \aap, 531, A29

\bibitem[{{K{\"u}ker} \& {R{\"u}diger}(1999)}]{kuk99}
{K{\"u}ker}, M. \& {R{\"u}diger}, G. 1999, \aap, 346, 922

\bibitem[{{Kuznetsov}(2011)}]{kuz10}
{Kuznetsov}, A.~A. 2011, \aap, 526, A161

\bibitem[{{Kuznetsov} \& {Tsap}(2007)}]{kuz07}
{Kuznetsov}, A.~A. \& {Tsap}, Y.~T. 2007, \solphys, 241, 127

\bibitem[{{Lamy} {et~al.}(2010){Lamy}, {Schippers}, {Zarka}, {Cecconi},
  {Arridge}, {Dougherty}, {Louarn}, {Andr{\'e}}, {Kurth}, {Mutel}, {Gurnett},
  \& {Coates}}]{lam10}
{Lamy}, L., {Schippers}, P., {Zarka}, P., {Cecconi}, B., {Arridge}, C.~S.,
  {Dougherty}, M.~K., {Louarn}, P., {Andr{\'e}}, N., {Kurth}, W.~S., {Mutel},
  R.~L., {Gurnett}, D.~A., \& {Coates}, A.~J. 2010, \grl, 37, L12104

\bibitem[{{Linsky} {et~al.}(1992){Linsky}, {Drake}, \& {Bastian}}]{lin92}
{Linsky}, J.~L., {Drake}, S.~A., \& {Bastian}, T.~S. 1992, \apj, 393, 341

\bibitem[{{Melrose} \& {Dulk}(1982)}]{mel82}
{Melrose}, D.~B. \& {Dulk}, G.~A. 1982, \apj, 259, 844

\bibitem[{{Mohanty} \& {Basri}(2003)}]{moh03}
{Mohanty}, S. \& {Basri}, G. 2003, \apj, 583, 451

\bibitem[{{Morin} {et~al.}(2010){Morin}, {Donati}, {Petit}, {Delfosse},
  {Forveille}, \& {Jardine}}]{mor10}
{Morin}, J., {Donati}, J.-F., {Petit}, P., {Delfosse}, X., {Forveille}, T., \&
  {Jardine}, M.~M. 2010, \mnras, 407, 2269

\bibitem[{{Mottez} {et~al.}(2010){Mottez}, {Hess}, \& {Zarka}}]{mot10}
{Mottez}, F., {Hess}, S., \& {Zarka}, P. 2010, \planss, 58, 1414

\bibitem[{{Osten} {et~al.}(2006){Osten}, {Hawley}, {Bastian}, \&
  {Reid}}]{ost06}
{Osten}, R.~A., {Hawley}, S.~L., {Bastian}, T.~S., \& {Reid}, I.~N. 2006, \apj,
  637, 518

\bibitem[{{Osten} {et~al.}(2009){Osten}, {Phan-Bao}, {Hawley}, {Reid}, \&
  {Ojha}}]{ost09}
{Osten}, R.~A., {Phan-Bao}, N., {Hawley}, S.~L., {Reid}, I.~N., \& {Ojha}, R.
  2009, \apj, 700, 1750

\bibitem[{{Phan-Bao} {et~al.}(2007){Phan-Bao}, {Osten}, {Lim}, {Mart{\'{\i}}n},
  \& {Ho}}]{pha07}
{Phan-Bao}, N., {Osten}, R.~A., {Lim}, J., {Mart{\'{\i}}n}, E.~L., \& {Ho},
  P.~T.~P. 2007, \apj, 658, 553

\bibitem[{{Ravi} {et~al.}(2011){Ravi}, {Hallinan}, {Hobbs}, \&
  {Champion}}]{rav11}
{Ravi}, V., {Hallinan}, G., {Hobbs}, G., \& {Champion}, D.~J. 2011, \apjl, 735,
  L2

\bibitem[{{Reiners} \& {Basri}(2007)}]{rei07}
{Reiners}, A. \& {Basri}, G. 2007, \apj, 656, 1121

\bibitem[{{Reiners} \& {Basri}(2010)}]{rei10}
---. 2010, \apj, 710, 924

\bibitem[{{Reiners} {et~al.}(2009){Reiners}, {Basri}, \& {Browning}}]{rei09}
{Reiners}, A., {Basri}, G., \& {Browning}, M. 2009, \apj, 692, 538

\bibitem[{{Russell} \& {Dougherty}(2010)}]{rus10}
{Russell}, C.~T. \& {Dougherty}, M.~K. 2010, \ssr, 152, 251

\bibitem[{{Schrijver}(2009)}]{sch09}
{Schrijver}, C.~J. 2009, \apjl, 699, L148

\bibitem[{{Treumann}(2006)}]{tre06}
{Treumann}, R.~A. 2006, \aapr, 13, 229

\bibitem[{{Wu} \& {Lee}(1979)}]{wu79}
{Wu}, C.~S. \& {Lee}, L.~C. 1979, \apj, 230, 621

\bibitem[{{Zarka}(1998)}]{zar98}
{Zarka}, P. 1998, \jgr, 103, 20159

\end{thebibliography}

\end{document}